\documentclass[twocolumn,showpacs,preprintnumbers,amsmath,amssymb,superscriptaddress,prl]{revtex4}
\usepackage{graphicx}
\usepackage{epstopdf}
\usepackage{dcolumn}
\usepackage{bm}
\usepackage[english]{babel}

\usepackage{caption}

\usepackage{amsmath,amssymb,bm,bbm}
\usepackage{graphicx}
\usepackage{multirow}
\usepackage{color}
\usepackage[normalem]{ulem}
\usepackage{tikz}

\usepackage{float}
\restylefloat{figure}

\usepackage{mathrsfs}

\newcommand{\ket}[1]{ | #1 \rangle }

\newcommand{\overlap}[2]{\langle #1 | #2 \rangle}
\newcommand{\elmx}[3]{\langle #1 | #2 | #3 \rangle}

\begin{document}

\title{First complete description of low-lying spectroscopy in $^{254}$No}

\author{Duy Duc Dao}
\author{Frédéric Nowacki}
\affiliation{Universit\'e de Strasbourg, CNRS, IPHC UMR7178, 23 rue du Loess, F-67000 Strasbourg, France}

\date{\today}

\begin{abstract} 
  In this work, we report the first complete shell-model description of low-lying structures of $^{254}$No. Employing the Kuo-Herling effective interaction, the calculations are performed within the Discrete Non-Orthogonal Shell Model recently implemented with the angular-momentum Variation After Projection applied on non-axial wavefunctions. Our calculations show a striking agreement with the experimentally known spectroscopy: the \textit{Yrast} band, the $8^{-}$ ($K=8$), $3^{+}$ ($K=3$) and $10^+$  ($K=10$) isomers together with associated $K$ bands. We reproduce the recently measured  Gallagher-Moszkowski splitting  between the $3^+$ ($K=3$) and $4^+$ ($K=4$) band heads. We predict the appearance of a second $0^+$ state, in excellent agreement with recent new observation. In addition, we systematically examine the ground state spectra, dipole and spectroscopic quadrupole moments of even-even, odd-even, odd-odd nuclei from $A=251$ to $A=256$, which are favourably reproduced. The present description of $^{254}$No shows the ability of the Shell Model framework to describe the low-lying properties for such an exotic nucleus at the frontier of modern experimental nuclear physics research. 
\end{abstract}


\maketitle

%
%
\noindent\textit{Introduction.}
In the search of unknown elements that delineates the border of the periodic table, superheavy nuclear elements, loosely defined with $Z\geq 102$, represent one of the most important regions at the forefront of contemporary experimental nuclear physics~\cite{Giu2019}. At the limit of charge and mass, the only way to study these exotic systems experimentally is to deal with single atoms due to low production rates and short nuclear lifetimes~\cite{Smits2024}. 
In the region of transfermium nuclei approaching the mass number $A\sim 250$, the $^{254}$No nucleus is a case of great interest that marks a breakthrough in superheavy nuclear structure studies~\cite{Reiter1999,Leino1999}. Compared to other neighbouring nuclei, its favourable experimental production cross section has allowed for detailed spectroscopy experiments from the ground state band properties up to high angular momentum~\cite{Ackermann2017}. In particular, the observation of high-$K$ isomers (where $K$ is the projection quantum number of the total angular momentum onto the symmetry axis) and associated $K$-bands give rise to a very complex level structure. The identification of these isomers and their decay patterns in this deformed nucleus has become a recurrent question for experimental investigations~\cite{Herzberg2006_Nature,Tandel2006_PhysRevLett.97.082502,CLARK2010_No254PLB,Hebberger2010EPJA_No254}. These studies reveal the appearance of several isomeric states, e.g. the long-lived $K^\pi=8^-$ isomer~\cite{Herzberg2006_Nature} or a possible $K^\pi=10^+$ state~\cite{CLARK2010_No254PLB,Clark2025_PhysRevC.111.034320}, 
as well as several non-\textit{Yrast} side bands, showing the richness and different facets that emerge in the structure of $^{254}$No. Recent investigations continue to provide new data, in particular, with the existence of a $K^\pi=4^+$ state interpreted as the Gallagher-Moszkowski partner of the $K^\pi=3^+$ and a second excited $0^+$ state which is observed for the first time~\cite{ForgePhD2023,No254PRL}. 

The rich spectroscopy information of $^{254}$No makes a strong case for theoretical investigations and testing nuclear models. Since the first observation of the rotational band in this nucleus~\cite{Reiter1999,Leino1999}, pioneering theoretical calculations were undertaken at the turn of the century in numerous studies under the Energy Density Functional (EDF) frameworks~\cite{Robledo2000_PhysRevLett.85.1198,DUGUET2001,Quentin2001,BENDER2003_No254,DELAROCHE2006,Zhang2012_PhysRevC.85.014324,Liu2012_PhysRevC.86.011301,Zhang2013_PhysRevC.87.054308,Shi2014_PhysRevC.89.034309}, dedicated to estimations of nuclear masses, deformation landscapes, fission-related properties and the pattern of single-particle spectra. Presently, non-relativistic and relativistic EDF models remain the unique microscopic theoretical tools to study superheavy nuclei~\cite{Bender2003_RevModPhys.75.121,Ring2005_VRETENAR2005,Robledo2014_GognyConstraints_PhysRevC.89.021303}. These theoretical descriptions center on the concept of intrinsic deformed mean fields derived from the EDFs within the Hartree-Fock-Bogoliubov (HFB) approximation. With the prevalence of the axial symmetry in many deformed nuclei beyond Lead~\cite{Walker1999_Nature}, microscopic EDFs models provide systematic estimations of high-$K$ isomers excitation energies and have been very useful to interpret their underlying structures in terms of quasi-particle couplings~\cite{Jolos_2011_Kisomers,Robledo_2024_Kisomers,Bonneau2024}. However, to study $K$-mixing effects due to beyond-mean-field correlations~\cite{Bender2008_EPJAS} and hence extract in particular the electromagnetic E2/M1 transitions of interest, it is necessary to perform configuration-mixing calculations with the restoration of the rotational symmetry. These calculations under the EDF frameworks still remain extremely difficult with only a few examples. For instance in~\cite{Heenen2016} the angular-momentum projection after variation (so-called PAV~\cite{RingSchuck1980}) was performed upon using a single HFB state for transfermium nuclei. In a more recent work~\cite{Egido2020_SHE_PhysRevLett.125.192504}, triaxial configuration-mixing calculations with the Gogny EDF were carried out in superheavy systems in the neutron magic number $N=184$ region.

 The nuclear Shell Model, despite its successful description of low-lying properties of light- and mid-mass nuclei~\cite{Caurier2005_RevModPhys.77.427,Otsuka2020_RevModPhys.92.015002,NOWACKI2021}, has never been applied to study superheavy systems. In principle, with a sufficiently large valence space composed of orbitals near the Fermi surface, by configuration mixing the Shell Model is capable to account for single-particle motions as well as low-lying nuclear collective excitations in the valence space. The lack of Shell Model configuration-mixing calculations in the superheavy region is mainly related to the difficulty to solve the Shell Model eigenvalue problem in large valence spaces by direct diagonalization where the number of basis states can reach $O(10^{29})$ for a superheavy nucleus as $^{254}$No, compared to the currently feasible dimension of order $O(10^{12})$ in medium-mass nuclei. Alternative approaches have been developed in numerous frameworks, e.g. the Monte-Carlo Shell Model~\cite{Otsuka-MCSM}, the Quasi-Vacua Shell Model (QVSM)~\cite{QVSM},  the Particle-Number Variation After Projection (PNVAP) of HFB wavefunctions combined with the Projected Generator Coordinate Method (PGCM)~\cite{TaurusSD,TaurusPF}, the recent Discrete Non-Orthogonal Shell Model (DNO-SM) and notably the famous Variation After Mean-field Projection In Realistic model spaces (VAMPIR) approach~\cite{SCHMID2004VAPHFB,Petrovici2024}. 

The common feature of these frameworks is based on the concept of symmetry breaking and restoration~\cite{Sheikh2021}. As nuclear states are characterized by "good" quantum numbers such as: angular momentum $J$, parity $\pi$, neutron and proton numbers $N,Z$, one can attempt to build sophisticated "intrinsic" states $\{\ket{\phi_i},i=1,2,...\}$ which deliberately break all possible symmetries in the system. The correlated nuclear state of good quantum numbers is then recovered by restoration of the broken symmetries through projection techniques and represented as a superposition
\begin{equation}
\ket{\psi^{\pi JM}_n} = \sum_{i,K}C^{\pi J}_n(i,K) 
\mathcal P^J_{MK}\:P^\pi\:P^N\:P^Z \ket{\phi_i},
\label{Psi}
\end{equation}
where $n$ labels physical states, $M,K$ the angular momentum projection quantum numbers in the laboratory and intrinsic frames, $C^{\pi J}_n(i,K)$ the mixing amplitudes. $\mathcal P^J_{MK},P^\pi,P^N,P^Z$ are the angular momentum, parity, neutron and proton projection operators~\cite{RingSchuck1980} defined as
\begin{equation}
\begin{aligned}
\mathcal P^J_{MK} &= \frac{2J+1}{8\pi^2} \int_0^{2\pi}d\alpha \int_0^{\pi}d\beta \int_0^{2\pi}d\gamma D^{J*}_{MK}(\Omega) \hat R(\Omega), \\
P^\pi &= \frac{1}{2}(1+ \pi\hat\Pi), \:
P^X = \frac{1}{2\pi}\int_0^{2\pi}d\varphi_X e^{i(\hat X - X)\varphi_X}.
\end{aligned}
\end{equation}
Here $\Omega=(\alpha,\beta,\gamma)$ represents the Euler angles, $D^{J}_{MK}(\Omega)$ the Wigner rotation matrix and $\hat R(\Omega)$ ($\hat\Pi$) the rotation (parity) operator. $X=N$ ($X=Z$) denotes the neutron (proton) number with the associated gauge angle $\varphi_X$. Starting from a two-body effective Hamiltonian $\hat H$, the most general way to apply the variational principle to the "trial" state~\eqref{Psi} is the double variation after projection (VAP) which consists in minimizing the projected energy
\begin{equation}
E^J_n = \frac{\elmx{\psi^{\pi JM}_n}{\hat H}{\psi^{\pi JM}_n}}{\overlap{\psi^{\pi JM}_\alpha}{\psi^{\pi JM}_n}}.
\label{EJprojected}
\end{equation}
with respect to the mixing amplitudes $C^{\pi J}_n(i,K)$ and the intrinsic states $\{\ket{\phi_i}\}$ simultaneously. The first variation determining $C^{\pi J}_n(i,K)$ leads to the standard Hill-Wheeler equations, with the Hamiltonian and norm matrix elements $H^{\pi J}_{iK,i'K'} = \elmx{\phi_i}{\hat H\mathcal P^J_{KK'}P^\pi P^N P^Z}{\phi_{i'}}$, $N^{\pi J}_{iK,i'K'} = \elmx{\phi_i}{\mathcal P^J_{KK'}P^\pi P^N P^Z}{\phi_{i'}}$, 
\begin{equation}
\sum_{i'K'} \big( H^{\pi J}_{iK,i'K'} - E^{\pi J}_\alpha N^{\pi J}_{iK,i'K'}\big) C^{\pi J}_\alpha(i',K') = 0
\label{HW}
\end{equation}
which can be transformed, through a diagonalization of the norm matrix $N^{\pi J}$, into a standard eigenvalue problem $\mathcal H^{\pi J} \:C^{\pi J} = E^{\pi J}\:C^{\pi J}$ with $\mathcal H^{\pi J} = [N^{\pi J}]^{-1/2}\: H^{\pi J} \: [N^{\pi J}]^{-1/2}$~\cite{Tomas2010_PhysRevC.81.064323,DNO2022}. The second variation determining $\ket{\phi_i}$ can be realized using the Thouless theorem~\cite{THOULESS1960225,MANG1975} where each intrinsic state is defined through a skew-symmetric $Z_{\mu\nu}$ and a arbitrarily fixed reference state $\ket{\phi_0}$
\begin{equation}
\ket{\phi_i} = \mathcal N e^{\frac 1 2\sum_{\mu\nu}Z_{\mu\nu}a^\dagger_\mu a^\dagger_\nu}\ket{\phi_0}.
\label{phi}
\end{equation}
This is the case of the VAMPIR approach~\cite{SCHMID2004VAPHFB} employing the most general HFB wavefunctions which are determined in a sequential manner, i.e. starting from the first $N$ minimized intrinsic states, one successively adds the $N+1$ intrinsic state and performs the second variation only for the last one $\ket{\phi_{N+1}}$. It should be noted that for example the PNVAP-PGCM framework uses a different strategy~\cite{TaurusSD,TaurusPF} keeping only the particle number in the second variation complemented with e.g. triaxial constraints. \\

Following the same line as~\cite{SCHMID2004VAPHFB} to define intrinsic states sequentially, we have recently implemented the DNO-SM within the double variation after projection framework (referred to as DNO-SM(VAP)) to tackle the secular Shell Model problem. Our approach is however different in several important aspects that are worth mentioned.
\begin{itemize}
\item[i)] Instead of using instrinsic HFB wavefunctions that break the neutron and proton numbers, we stick to Slater-type determinants;
\item[ii)] We are able to treat the most general non-axial wavefunctions that break the parity symmetry, the rotational symmetry and point group symmetries~\cite{DobacI2000_PhysRevC.62.014310,DobacII2000_PhysRevC.62.014311}, therefore do not assume particular nuclear shapes in the intrinsic states ;
\item[iii)] Correlations (e.g. of triaxial quadrupole, octupole or pairing types) allowed in the valence space are incorporated through the configuration mixing of VAP non-axial wavefunctions without breaking the particle numbers ;
\item[iv)] With the angular-momentum variation after projection, we are also able to treat even-even, odd-even and odd-odd nuclei on an equal footing as other frameworks using HFB wavefunctions~\cite{TaurusPF,SCHMID2004VAPHFB}.
\end{itemize}
The first feature i) is essential to make calculations much less costly, especially for superheavy systems. Indeed, as Slater-type determinants preserve the neutron and proton numbers, it is not necessary to perform the restoration of the particle number. The nuclear wavefunction~\eqref{Psi} is simplified into 
\begin{equation}
\ket{\psi^{\pi JM}_n} = \sum_{i,K}C^{\pi J}_n(i,K) 
\mathcal P^J_{MK}\:P^\pi \ket{\phi_i}.
\label{Slater}
\end{equation} 
A fivefold integration is therefore reduced to a threefold integration, which greatly simplify the evaluation of Hamiltonian matrix elements. Secondly, it is always desirable to be able to incorporate the most general non-axial degrees of freedom present in the valence space. For example in transfermium nuclei around $^{254}$No and in superheavy systems towards the island of stability close to $N=184$, non-axial degrees of freedom such as triaxial octupole~\cite{Ring2024_PhysRevLett.133.022501} and triaxial quadrupole deformations~\cite{Egido2020_SHE_PhysRevLett.125.192504} have been realized to be important. Thirdly, it is usually assumed that the description of pairing correlations require the use of HFB wavefunctions~\cite{RingSchuck1980,SCHMID2004VAPHFB,Sheikh2021}. From the formal point of view, the solutions are independent of the (simple or sophisticated) many-body basis used for diagonalization of the given Hamiltonian. We mention two examples to show the efficiency of Slater-type determinants in the double variation after projection. With the non-trivial KB3G interaction as~\cite{TaurusPF}, compared to the exact diagonalization, we recover the ground state energy within $20$ keV using $36$ Slater states for $^{48}$Cr, a typical deformed nucleus exhibiting the backbending behavior due to pairing correlations~\cite{Caurier2005_RevModPhys.77.427}. In a much larger valence space for $^{62}$Cr (see~\cite{62Cr_Nature} for details), the exact Shell Model ground state energy is $-214.138$ MeV from the diagonalization in the complete $M$-scheme basis of $2\times 10^9$ spherical harmonic oscillator Slater states. The PAV calculation, using $711$ deformed Hartree-Fock states, give the ground state binding energy $-213.978$ MeV. Our VAP calculation yields $-214.048$ MeV with $73$ Slater determinants. Finally, it is worth noting that although the double variation after projection approach has existed for a long time, so far there is no calculations using such framework applied to general non-axial wavefunctions 
for studies of superheavy nuclei. The most recent applications of the VAMPIR framework are presently restricted to the axially symmetric and time-reversal invariant HFB wavefunctions~\cite{Petrovici2024} for nuclei of mass $A < 100$.

Here we report the first complete description of the low-lying structures of $^{254}$No within the Shell Model framework using the DNO-SM(VAP) implementation. Employing the Kuo-Herling effective interaction designed more than twenty year ago, we first perform a systematic comparison of dipole magnetic and spectroscopic quadrupole moments as well as associated \textit{Yrast} spectra to the available experimental data for different situations of even-even, odd-even and odd-odd nuclei of masses ranging from $A=251$ to $A=256$ from which we extract effective charges relevant for this mass region. Secondly, for $^{254}$No we are able to reproduce the spectra of various isomeric states and associated $K$-band developed on top of them: $K^\pi=8^-,3^+,10^+$ at the same time as the ground state rotational band. In particular, we provide the prediction of a new $K^\pi=4^+$ state recently reported in~\cite{ForgePhD2023} which is identified as the Gallagher-Moszkowski partner of the known $K^\pi=3^+$ state. We also predict the existence of a second low-lying $0^+$ state which has been evidenced for the first time in $^{254}$No~\cite{ForgePhD2023}. \\

\begin{center}
\begin{figure}[H]  
    \vspace{-4.3em}
    \includegraphics[scale=0.258]{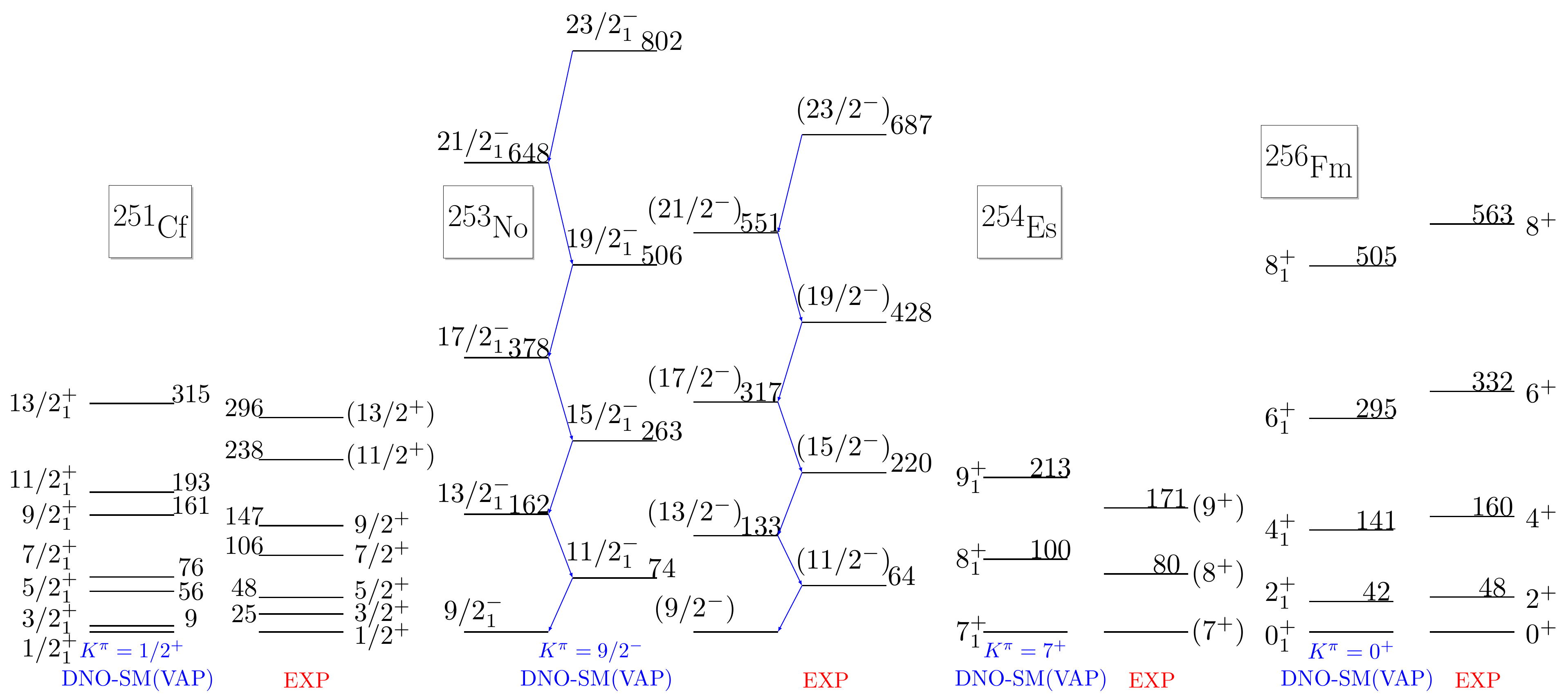}
    \caption{\textit{Yrast} rotational band spectra from DNO-SM(VAP) calculations using 1 Slater state in comparison with the experimental data (EXP) taken from~\cite{NDSA251,NDSA253,NDSA254,NDSA256}. In parentheses are tentative experimental spin-parity assignments. Excitation energies are given in keV.}
    \label{SM_Yrast}
\end{figure}
\end{center}

\noindent\textit{Results and discussion.}
The effective Kuo-Herling interaction and the associated valence space which we use contains $0h_{9/2},0i_{13/2},1f_{7/2},1f_{5/2},2p_{3/2},2p_{1/2}$ orbitals for protons and $0i_{11/2},0j_{15/2},1g_{9/2},1g_{7/2},2d_{5/2},2d_{3/2},3s_{1/2}$ for neutrons on top of the $^{208}$Pb core. Its matrix elements were obtained from realistic Hamada-Johnston potential~\cite{KUOHERLING1972,Warburton1991_KH_PhysRevC.43.602} and later minor phenomenological proton-proton monopole constraints were incorporated by E. Caurier and F. Nowacki~\cite{Hauschild2001_PhysRevLett.87.072501,Caurier2003_PhysRevC.67.054310} using data around a $^{208}$Pb region. Thus no adaptation to the mass region under consideration was introduced.
%
%
To test the validity of the Kuo-Herling interaction, we first consider systematic calculations of A$\sim$ 250 nuclei. The binding energies,
the magnetic dipole moment $\mu$ and the spectroscopic quadrupole moment $Q_s$ are shown in Table~\ref{QMu}. For the determination of effective charges, we perform a fit based on the sample of nuclei in Table~\ref{QMu} and obtain the values $e_p=1.72$, $e_n=0.75$ for protons and neutrons respectively. For the computation of magnetic moments, bare gyromagnetic factors are used. These calculations are performed using $1$ Slater state in both PAV and VAP schemes. In the former, we use the cranking method as in~\cite{DNO2022} with the constraint on the expectation value of the third angular moment component $\langle J_z\rangle = J_{\mathrm gs}$. Under these conditions, we determine the ground state of $^{253}$No, $^{253,254,255}$Es and $^{249,251,253}$Cf isotopes whose magnetic dipole and quadrupole moments have recently been measured by~\cite{Raeder2018No253,Nothhelfer2022_PhysRevC.105.L021302,Weber2023Cf}.
The ground state bindings obtained in the VAP calculations are systematically lower than the PAV values, which is expected from the variational principle point of view. It is observed that the energy difference between VAP and PAV calculations is of order $\sim 1$ MeV in $^{253}$No and $^{254,255}$Es (whereas it is about $\sim 400$ keV in other odd cases) where the PAV quadrupole and dipole moments are of opposite signs or considerably lower than the experimental values. This underestimation points to the important role of correlations, hence the superiority of the VAP scheme to systematically capture them with respect to the PAV scheme, making the theoretical magnetic and quadrupole moments converge towards the order of the experimental values with the correct sign.
We observe a remarkable agreement with the experimental values for all considered odd nuclei. Note that the ground-state spin of the odd-odd $^{254}$Es is tentatively assigned $J=7$ and its parity is not known~\cite{Nothhelfer2022_PhysRevC.105.L021302}. Our calculation predicts its ground state to be $J^\pi=7^+$, with an almost pure $K=7$ of $99.51\%$ in its wave function and a residual mixing of $0.49\%$ $K=6$ component. Overall, both calculations PAV and VAP agree with experimental values within the quoted uncertainties.  The reproduction of the electromagnetic moments therefore indicates a very good spectroscopic quality of the Kuo-Herling interaction in this heavy-actinide mass region. In Fig.~\ref{SM_Yrast} are the \textit{Yrast} bands obtained from the same DNO-SM(VAP) calculations with 1 Slater state in comparison with data. Along with the reproduction of the magnetic and quadrupole moments, the agreement between theory and experiment is excellent for both the spin-parity predictions and energy level orderings up to very good precision in the range of $\sim 10-50$ keV. 
%
%
%
%

Given this successful description in the surrounding nuclei, we investigate the specific case of $^{254}$No. The convergence of the band heads $0^+_1,3^+_1$ and $8^{-}_2$ is given in Fig.~\ref{Nq254No} where the inset shows the comparison of the ground state binding computed with the projection after variation of the Hartree-Fock minimum (PAV(1)), the previous DNO-SM calculation~\cite{DNO2022} with $15$ $(\beta,\gamma)$-constrained Hartree-Fock states (PAV(15)) and the present variation after projection calculation with 1 Slater state (VAP(1)). Here $\beta$ is the quadrupole deformation parameter and $\gamma$, varying from $0^\circ$ (prolate shape) to $60^\circ$ (oblate shape), measures the possible deviations from the axial symmetry. We note that with $1$ Slater state, the obtained binding energy in the VAP(1) calculation is already lower than 15 constrained Hartree-Fock states in the PAV scheme. Altogether the relative energy gains for the various states are not the same but rather depend on the structure of the bandheads themselves and the final VAP energy gains (in MeV) with respect to the PAV(15) calculation are $0.77$, $0.20$ and $0.84$ for $0^+_1$, $3^+_1$ and $8^-_2$ bandheads respectively.
Fig.~\ref{PES} shows the potential energy surface obtained from $(\beta,\gamma)$-constrained Hartree-Fock states and the mass quadrupole effective charges as defined in~\cite{DNO2022}. With the newly determined effective charges from Table~\ref{QMu}, the potential energy surface shows a single energy minimum lying on the prolate axial axis at $\beta \sim$ 0.25. The latter value is indeed compatible with different microscopic non-relativistic and relativistic mean-field predictions~\cite{Reiter1999,DUGUET2001,Robledo2000_PhysRevLett.85.1198,Quentin2001,Ring2024_PhysRevLett.133.022501} in the range $\beta \sim 0.25-0.3$.
The triaxial $(\beta,\gamma)$-configuration-mixing wavefunction for the ground state is also illustrated in Fig.~\ref{PES} where the orange circles show the various components of the state in the $(\beta,\gamma)$ plane. 
The ground state is found to be strongly dominated by $\gamma \sim 0^\circ$ configurations centered around the Hartree-Fock minimum, implying a predominant axially deformed structure.
%
%
%

\begin{figure}[H]
    \vspace{-1.4em}
    \includegraphics[scale=0.36]{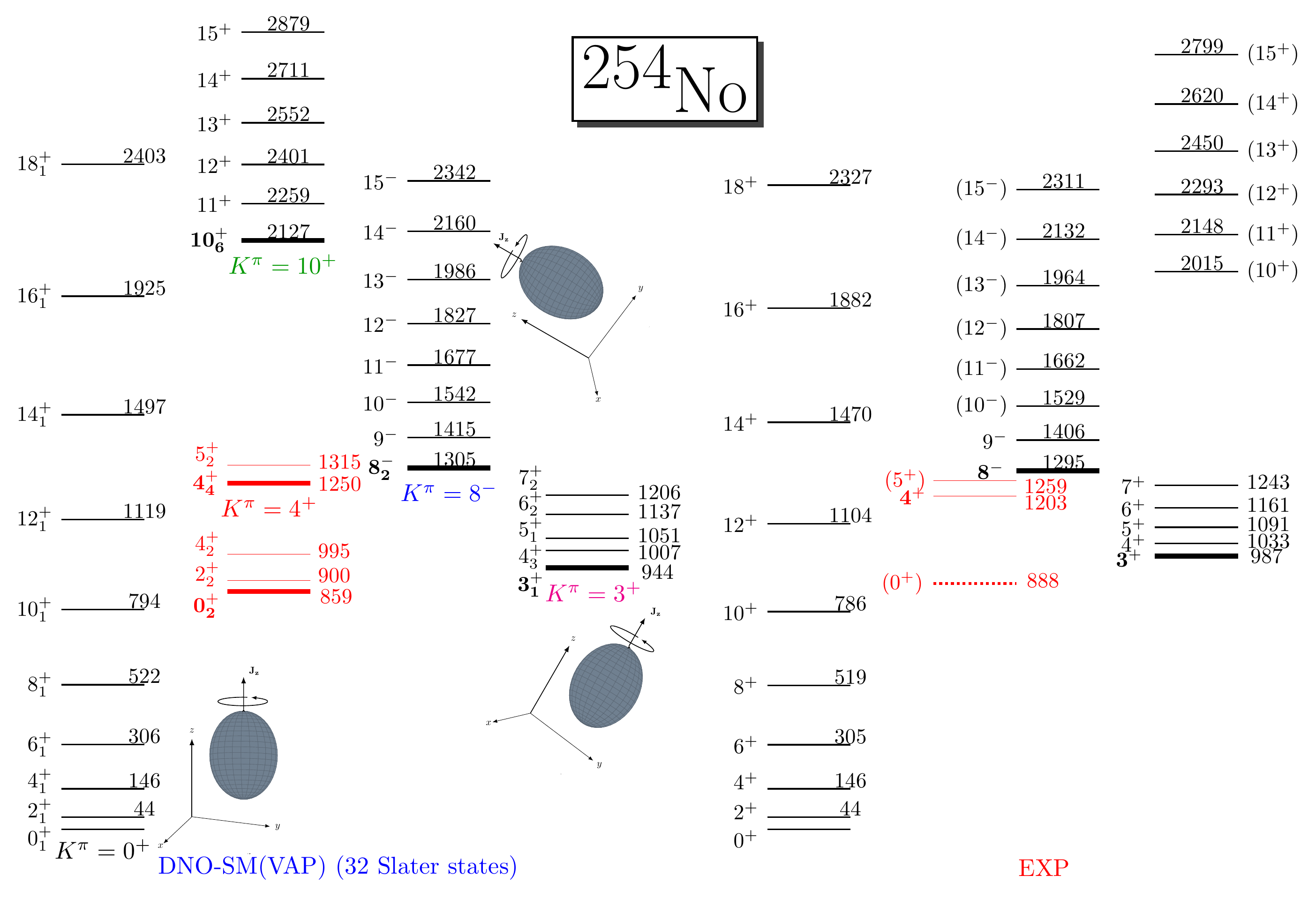}
    \caption{Spectra of \textit{Yrast} and excited bands from DNO-SM(VAP) calculations in comparison with the experimental data (EXP) taken from~\cite{Herzberg2006_Nature,CLARK2010_No254PLB,Hebberger2010EPJA_No254,ForgePhD2023,Clark2025_PhysRevC.111.034320}. The red levels are the newly proposed states from the recent study in~\cite{ForgePhD2023,No254PRL}. Tentative experimental spin-parity assignments are given between parentheses. Excitation energies are in keV. $K$ denotes the total angular momentum projection quantum number. The $K$-band heads are given in bold lines with the associated $K$ component which amounts up to $\sim 95\%$, probing the axial symmetry preservation in the correlated wavefunctions. The ellipsoids, with different orientations (pointing to the appearance of different $K$ bands), thus illustrate the analogy to a classic rotating axial rigid rotor (cf. pages 475-476~\cite{RingSchuck1980}).\label{SM_254No}}
\end{figure}

Fig.~\ref{SM_254No} presents the resulting spectra, for $^{254}$No, obtained from the DNO-SM(VAP) calculations including up to $32$ Slater states. The agreement with the experimental data for various excited band structures is striking with a one to one correspondence of all the states within  100 keV. An analysis of the microscopic correlated wave functions shows the $3^+_1$, $4^+_4$, $8^-_2$ and  $10^+_6$  states have a predominant component (up to $\sim 95\%$) $K=3$, $K=4$, $K=8$ and $K=10$  quantum numbers respectively and therefore are identified as the experimental bandheads. The K conservation 
points to the purity of the axial symmetry from the ground state to high angular momenta, which is reminiscent of an axial rigid rotor picture (as illustrated by the ellipsoids with different "$K$" orientations)~\cite{RingSchuck1980}. The rotational \textit{Yrast} band is perfectly reproduced up to spin 18$^+$ and the calculated B(E2) for the 2$^+_1$ to 0$^+_1$ transition amounts to 28666 e$^2$.fm$^4$. This value is similar to the ones reported for $^{248}$Cm and $^{250}$Cf N=152 isotones nearby~\cite{PRITYCHENKO20161}. In Table~\ref{E2M1} we collect the theoretical electromagnetic E2/M1 transition rates
for the low-lying states up to spin $J=4$. The collective structures are identified by  large intra-band B(E2) transitions. The decay-out of various bandheads (notably 4$^+_4 (K=4)$ to  3$^+_1 (K=3)$ and to \textit{Yrast} states) proceeds through transitions originating from small $K$-quantum number admixtures in the correlated wavefunctions. For example the $K=3$ and $K=2$
components of the 4$^+_4$ state amount to $4~\%$ and $0.013~\%$ respectively. Consequently the obtained  B(E2, $4^+_4 \rightarrow 3^+_1$) transition  is $\sim 10^{7}$ larger than the  B(E2, $4^+_4 \rightarrow 2^+_1$) and explain the recently observed gamma-ray transition between the two Gallagher-Moszkowski 4$^+_4 (K=4)$ and 3$^+_1 (K=3)$ partners~\cite{No254PRL}.  We also  probe the collective features of the $K^\pi=3^+$ side band  by looking at the ratio between B(E2, $7^+_2 \rightarrow 5^+_1$) and  B(M1, $7^+_2 \rightarrow 6^+_2$). The calculations produce B(E2, $7^+_2 \rightarrow 5^+_1$) =   23971 e$^2$.fm$^4$ and  B(M1, $7^+_2 \rightarrow 6^+_2$) =  0.228 $\mu_N^2$ leading to a branching ratio of 0.94 to be compared to $\sim$ 1.1(4) experimentally determined in~\cite{Tandel2006_PhysRevLett.97.082502,No254PRL}.

Table~\ref{OccSM} presents the spherical occupancies of various band heads in the spectrum. The $3^+_1$ $(K=3)$ and $4^+_4$ $(K=4)$ band heads exhibit very similar occupancies with a proton excitation with respect to the ground state. The energy difference between these two states was experimentally observed at $216$ keV and interpreted following the Gallagher-Moszkowski splitting~\cite{No254PRL}. Indeed, from Fig.~\ref{SM_254No}, the calculated splitting is very close to the experimental value. The $8^-_2 (K=8)$ isomer essentially originates, with respect to the ground state, from a neutron-neutron coupling $[\nu g9/2\otimes\nu j15/2]$ with some slight residual mixing of the other orbitals. The $10^+_6 (K=10)$ isomer structure mainly stems from a $4$-particles recoupling $[\pi h9/2\otimes \pi i13/2] \times [\nu j15/2\otimes \nu g7/2]$.  Finally the recently observed $0^+_2$ candidate~\cite{No254PRL} at $888$ keV has its theoretical counterpart lying at $859$ keV. Inspection of occupancies in Table~\ref{OccSM} reveals the proton nature of this state. \\


\noindent\textit{Conclusion}. In summary, we have presented the first complete shell-model description of low-lying structures of $^{254}$No with an implementation of the Discrete Non-Orthogonal Shell Model using the angular-momentum variation after projection applied on non-axial wavefunctions. The results obtained in this work show that the VAP is able to capture important correlations with respect to the PAV scheme. The description of exotic nuclei in the mass region around $^{254}$No and their spectroscopic systematics appears to be excellent. In the typical case of $^{254}$No, we reach a remarkable reproduction of its many facets: the rotational band and all known excited  structures, including in particular the newly identified second 0$^+$ state and the $K=4$ ($4^+$, $5^+$) structure~\cite{ForgePhD2023,No254PRL}. Our results allow for a microscopic description of very heavy and very deformed nuclei within the shell-model framework, and calls for broader systematic studies in the region. Lastly, the present description proposes a benchmark to be confronted to modern shell-model interactions from ab-initio Valence-Space IM-SRG approaches~\cite{Stroberg2019_VSIMSRG}. 
%
%
\onecolumngrid
\begin{center}
\begin{table}[H]
\centering
\scalebox{1.1}{\begin{tabular}{ccccccccccccc}
\hline\hline
\multicolumn{2}{c}{$J^\pi_{gs}$} & \multicolumn{2}{c}{E (MeV)}  && \multicolumn{3}{c}{$\mu (\mu_N)$} && \multicolumn{3}{c}{$Q_s$ (eb)} \\ \cline{3-4}  \cline{6-8} \cline{10-12}
&    &  PAV & VAP && PAV & VAP & EXP  &&  PAV & VAP & EXP \\ 
\hline
$^{253}$No & $9/2^-$ & $-241.818$ & $-242.816$ &&  $+0.591$ & $-0.493$ & $-0.527(33)(75)$ &&   $-3.5$   & $+7.2$ & $+5.9(1.4)(0.9)$ \\
$^{253}$Cf & $7/2^+$ & $-253.402$ & $-253.818$ && $-0.677$ & $-0.556$ & $-0.731(35)$ && $+5.77$ & $+5.78$ & $+5.53(51)$\\
$^{251}$Cf & $1/2^+$ & $-241.321$  & $-241.724$ &&  $-0.727$ & $-0.610$ & $-0.571(24)$ &&   -    &  - & - \\
$^{249}$Cf & $9/2^-$ & $-229.021$  & $-229.381$    &&  $-0.480$    & $-0.461$ & $-0.395(17)$ &&  $+6.62$     & $+6.63$ & $+6.27(33)$ \\
$^{255}$Es & $7/2^{+}$ & $-263.512$  &  $-264.695$ &&  $-1.10$   & $+3.94$ & $+4.14(10)$  &&  $+6.0$    & $+5.8$ & $+5.1(1.7)$\\
$^{254}$Es & $7^{+}$ & $-257.492$ & $-258.441$ &&  $+0.778$ &  $+3.36$ & $+3.42(7)$  &&  $+1.8$ & $+8.4$ & $+9.6(1.2)$\\
$^{253}$Es & $7/2^{+}$ &  $-251.837$  &  $-252.280$    &&  $+3.63$   & $+3.93$ & $+4.10(7)$ &&  $+5.87$     & $+5.9$ & $+6.7(8)$ \\
\hline 
$^{256}$Fm & $0^{+}$ &  $-268.999$  &  $-269.717$    &&  $+0.87$   & $+0.89$ & - &&  $-3.57$     & $-3.60$ & - \\
\multirow{1}{*}{$^{254}$No} & \multirow{1}{*}{$0^{+}$} &  $-249.568$  &  $-250.187$    &&  $+0.87$   & $+0.91$ & - &&  $-3.78$ & $-3.75$ & - \\
\hline\hline
\end{tabular}}
\caption{Spectroscopic quadrupole ($Q_s$ in eb unit) and magnetic dipole ($\mu$ in nuclear magneton $\mu_N$ unit) moments of the ground state of spin and parity $J_{\mathrm gs},\pi$ obtained from DNO-SM calculations with 1 cranked Hartree-Fock state (PAV) in comparison with 1 variation after projection Slater state (VAP) and the experimental data (EXP) from Refs.~\cite{Weber2023Cf,Raeder2018No253}. For $^{256}$Fm,$^{254}$No, we show the predicted quadrupole and magnetic moments of the first $2^+$ state. Effective charges are $e_p=1.72,e_n=0.75$ and bare values of the gyromagnetic factors are used. The absolute ground state binding ($E$ in MeV) is also shown for comparison between two calculations.}
\label{QMu}
\end{table}
\end{center}

\pagebreak


\begin{minipage}{0.48\textwidth}
\begin{figure}[H]
    \includegraphics[scale=1.6]{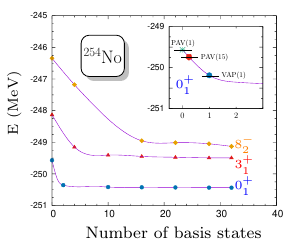}
    \caption{Convergence of the ground state $0^+_1$ and the band heads $3^+_1$, $8^-_2$ energies in $^{254}$No with the number of basis states.\label{Nq254No}}
\end{figure}
\end{minipage}
\hfill
\begin{minipage}{0.48\textwidth}
\begin{figure}[H]
    \includegraphics[scale=0.37]{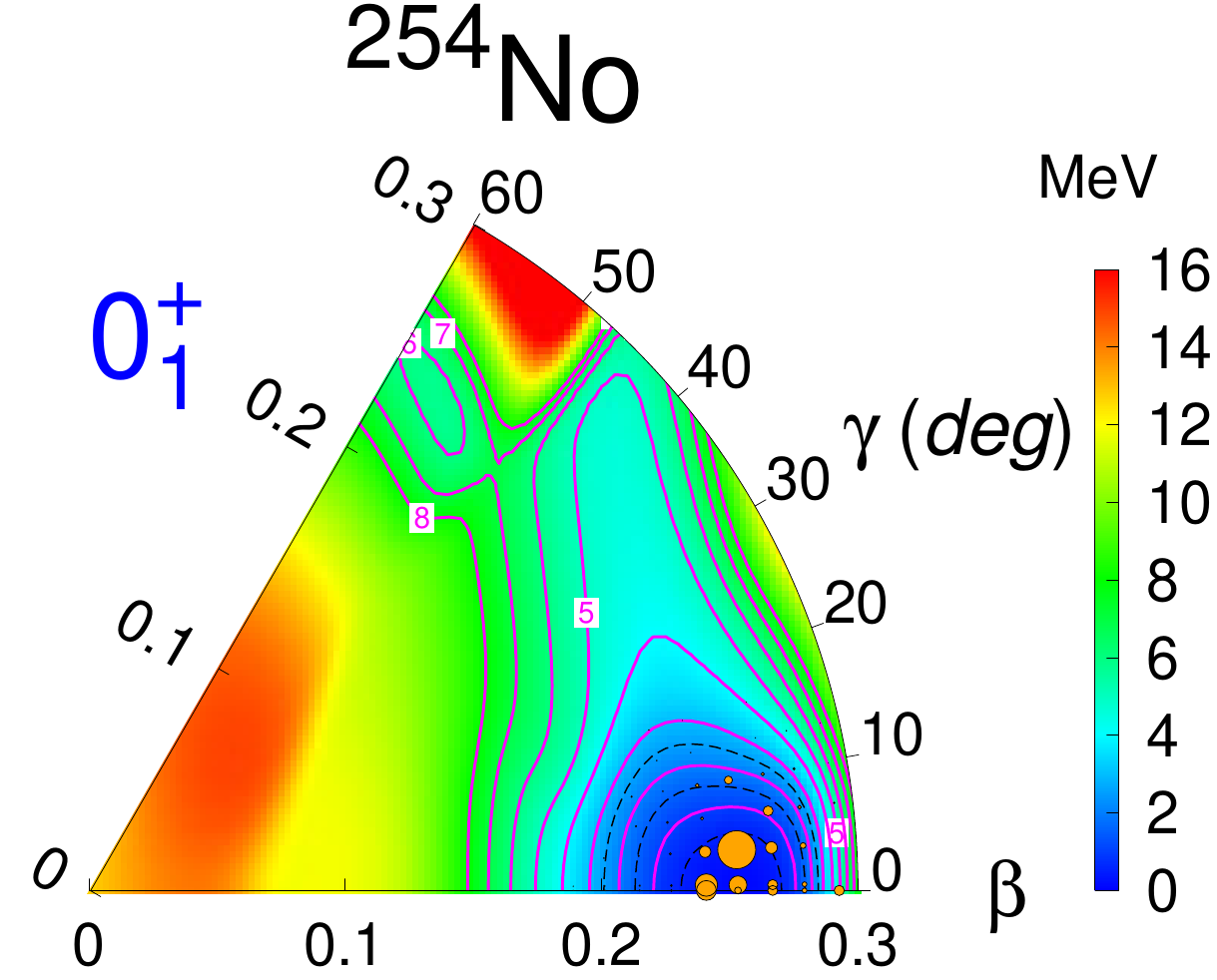}
    \caption{Ground state structure of $^{254}$No in the $(\beta,\gamma)$ potential energy surface. The area of orange circles is directly proportional to the normalized probability to find a deformation $(\beta,\gamma)$ in the corresponding state.\label{PES}}
\end{figure}
\end{minipage}

\pagebreak

\twocolumngrid

\begin{table*}
      \begin{tabular}{rlcc} \hline\hline
      $J_i^\pi \to$ & $J_f^\pi$   & $B(E2)$ (e$^2$.fm$^{4}$)  &   $B(M1)$ ($\mu_N^2$)\\
      \hline
      $2_1^+ \to$ & $0_1^+$ & 28666                  &     -    \\
      $4_1^+ \to$ & $2_1^+$ & 41021                  &     -    \\
      $0_2^+ \to$ & $2_1^+$ & 98.236                 &     -    \\
      $2_2^+ \to$ & $0_2^+$ & 25327                  &     -    \\
                  & $4_1^+$ & 60.738                 &     -     \\  
                  & $2_1^+$ & 25.227                 & 3.980$\times 10^{-6}$ \\
      $3_1^+ \to$ & $2_2^+$ & 2.5959$\times 10^{-2}$ & 8.020$\times 10^{-6}$ \\
                  & $4_1^+$ & 2.7359$\times 10^{-4}$ & 2.957$\times 10^{-7}$ \\
                  & $2_1^+$ & 9.6147$\times 10^{-4}$ & 6.050$\times 10^{-6}$ \\      
      $4_2^+ \to$ & $2_2^+$ & 36147                  &     -     \\
                  & $3_1^+$ & 5.4204$\times 10^{-1}$ & 4.200$\times 10^{-7}$  \\ 
                  & $4_1^+$ & 23.783                 & 8.360$\times 10^{-6}$  \\
                  & $2_1^+$ & 18.569                 &    -       \\      
      $4_3^+ \to$ & $3_1^+$ & 45803                  & 1.3852$\times 10^{-1}$ \\
                  & $6_1^+$ & 6.3712$\times 10^{-4}$ &  -  \\
                  & $4_1^+$ & 6.5482$\times 10^{-3}$ &  3.634$\times 10^{-5}$  \\
                  & $2_1^+$ & 8.2100$\times 10^{-6}$ &  -   \\      
      $4_4^+ \to$ & $3_1^+$ & 1.1354$\times 10^{3}$  & 1.977$\times 10^{-3}$ \\         
                  & $4_3^+$ & 9.6847$\times 10^{2}$  & 1.737$\times 10^{-2}$  \\
                  & $2_2^+$ & 4.438$\times 10^{-4}$  &    -       \\      
                  & $2_1^+$ & 1.398$\times 10^{-4}$  &    -       \\      

      \hline\hline          
                  \end{tabular}  
   \caption{\label{E2M1}  Intra-band and inter-band $B(E2)$ and $B(M1)$ transitions from the initial state $J^\pi_i$ to the final state(s) $J^\pi_f$ obtained for several low-lying states of the \textit{Yrast} rotational band, the excited $0^+_2$ band and the $3^+_1 (K=3),\:4^+_4 (K=4)$ bands in Fig.~\ref{SM_254No}.}
\end{table*}

\begin{table}[H]
\centering
\scalebox{0.9}{\begin{tabular}{*{8}c}
            \hline\hline
            \textcolor{blue}{Proton orbits} & $0h_{9/2}$ & $0i_{13/2}$ & $1f_{7/2}$ & $1f_{5/2}$ & $2p_{3/2}$ & $2p_{1/2}$ & \\
            \cline{2-8}
            $0^+_1$ & \textcolor{blue}{$6.03$} & \textcolor{blue}{$7.75$} & \textcolor{blue}{$3.43$} & $1.49$ & $0.77$ & $0.52$ & \\
            $0^+_2$ & \textcolor{blue}{$7.08$} & \textcolor{blue}{$7.91$} & \textcolor{blue}{$3.23$} & $0.89$ & $0.69$ & $0.21$ & \\
            $3^{+}_1 (K=3)$ & \textcolor{blue}{$6.47$} & \textcolor{blue}{$7.98$} & \textcolor{blue}{$3.34$} & $1.15$ & $0.72$ & $0.34$ \\
            $4^{+}_4 (K=4)$ & \textcolor{blue}{$6.50$} & \textcolor{blue}{$7.83$} & \textcolor{blue}{$3.41$} & $1.18$ & $0.72$ & $0.36$ \\
            $8^{-}_2 (K=8)$ & \textcolor{blue}{$6.48$} & \textcolor{blue}{$7.90$} & \textcolor{blue}{$3.36$} & $1.19$ & $0.70$ & $0.37$ \\
            $10^{+}_6 (K=10)$ & \textcolor{blue}{$6.55$} & \textcolor{blue}{$7.03$} & \textcolor{blue}{$3.48$} & $1.56$ & $0.79$ & $0.58$ \\            
            \hline
            \textcolor{red}{Neutron orbits} & $0i_{11/2}$ & $0j_{15/2}$ & $1g_{9/2}$ & $1g_{7/2}$ & $2d_{5/2}$ & $2d_{3/2}$ & $3s_{1/2}$ \\
            \cline{2-8}
            $0^+_1$ & \textcolor{red}{$7.30$} & \textcolor{red}{$9.91$} & \textcolor{red}{$5.43$} & $1.00$ & $1.09$ & $0.84$ & $0.43$ \\
            $0^+_2$ & \textcolor{red}{$7.36$} & \textcolor{red}{$9.95$} & \textcolor{red}{$5.45$} & $0.96$ & $1.05$ & $0.80$ & $0.42$ \\
            $3^+_1 (K=3)$ & \textcolor{red}{$7.32$} & \textcolor{red}{$9.94$} & \textcolor{red}{$5.46$} & $0.97$ & $1.07$ & $0.81$ & $0.42$ \\
            $4^+_4 (K=4)$ & \textcolor{red}{$7.34$} & \textcolor{red}{$9.79$} & \textcolor{red}{$5.48$} & $1.03$ & $1.11$ & $0.81$ & $0.44$ \\
            $8^-_2 (K=8)$ & \textcolor{red}{$7.42$} & \textcolor{red}{$9.00$} & \textcolor{red}{$6.30$} & $0.98$ & $1.06$ & $0.81$ & $0.43$ \\
            $10^+_6 (K=10)$ & \textcolor{red}{$7.23$} & \textcolor{red}{$8.89$} & \textcolor{red}{$5.72$} & $1.40$ & $1.34$ & $0.97$ & $0.45$ \\
            \hline\hline
        \end{tabular}}
        \caption{Spherical occupancies of the ground state $0^+_1$, the second $0^+_2$, the long and short-lived isomers $8^-_2,3^+_1$, and the $4^+_4$, $10^+_6$ obtained from the DNO-SM(VAP) calculations shown in Fig.~\ref{SM_254No}.}
        \label{OccSM}
\end{table}

\bibliography{main}
\end{document}